\documentclass[useAMS,usenatbib]{mn2e}

\pdfoutput=1

\usepackage{lscape}
\usepackage{graphicx,times}
\usepackage{float}
\newcommand{\lesssim}{\mbox{{\raisebox{-0.4ex}{$\stackrel{<}{{\scriptstyle\sim}}$}}}}

\title[Cygnus A: slow jet speeds and precession]{Multiwavelength study of
  Cygnus\,A I. Precession and slow jet speeds from radio observations} 
\author[K.C. Steenbrugge and
  K. M. Blundell]{Katrien
  C. Steenbrugge$^{1}$\thanks{E-mail:kcs@astro.ox.ac.uk} and Katherine
  M. Blundell$^{2}$\\ 
$^{1}$St John's College Research Centre, University of Oxford, St
  John's College, Oxford, OX1 3JP, UK\\ 
$^{2}$University of Oxford, Department of Physics, Keble Road, Oxford, OX1 3RH, UK}
\begin{document}

\date{Accepted . Received }

\pagerange{\pageref{firstpage}--\pageref{lastpage}} \pubyear{2002}

\maketitle

\label{firstpage}

\begin{abstract}
  We study the jet and counterjet of the powerful classical double
  FRII radio galaxy Cygnus\,A as seen in the 5, 8 and 15-GHz radio
  bands using the highest spatial resolution and signal-to-noise
  archival data available.  We demonstrate that the trace of the radio
  knots that delineate the jet and counterjet deviates from a straight
  line and that the inner parts can be satisfactorily fitted with the
  precession model of Hjellming \& Johnston. The parameter values of the
  precession model fits are all plausible although the jet speed is
  rather low (\lesssim 0.5\,$c$) but, on investigation, found to be
  consistent with a number of other independent estimates of the jet speed in
  Cygnus\,A. We compare the masses and precession periods for sources
  with known precession and find that for the small number of AGN with
  precessing jets the precession periods are significantly longer than
  those for microquasars.
\end{abstract}


\section{Introduction}\label{sect:intro}

\begin{figure}
\begin{center}
 \resizebox{\hsize}{!}{\includegraphics[angle=0]{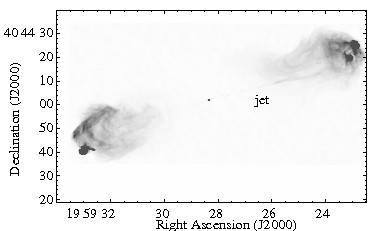}}
\caption{ The image of Cygnus\,A at 5 GHz made by Chris Carilli
   using all four VLA configurations and first published in
  \citet{carilli91}, showing the well known jet on the western,
  approaching side. Both lobes are brighter north of the jet
  trajectory. \label{fig:5_full}}
\end{center}
\end{figure}

Cygnus\,A (3C\,405), at a redshift of 0.05607 \citep{owen97}, is the
closest powerful FR\,II classical double radio galaxy (see
Fig.~\ref{fig:5_full}). At this distance, and for a Hubble constant of
73 km s$^{-1}$ Mpc$^{-1}$ (we also assume that $\Omega_{\rm
M}=0.3$ and $\Omega_{\Lambda} = 0.7$), 1$^{\prime\prime}$ corresponds
to 0.9551~kpc, giving good physical resolution on its different
structures and so Cygnus\,A has been well studied in the radio
\citep[e.g.][]{carilli91}.  

In the nuclear region, a secondary point source has been identified by
\citet{canalizo03} from a Keck\,II adaptive optics near-infrared
image. The most likely explanation is that this is a remnant nucleus,
due to a minor merger. In addition, redshifted H 21 cm absorption
lines \citep{conway95} and redshifted H$_2$ emission lines
\citep{bellamy04} are detected against the radio core, probably
indicating the inflow of gas. A possible explanation is the infall of
a giant molecular cloud, with the resultant radial gas motions
fuelling the active galactic nucleus (AGN) and its jets and lobes
\citep{bellamy04}. There is therefore quite some evidence for merger
activity in this galaxy \citep{canalizo03}, although no large-scale
structural disruption seems to have occurred \citep{owen97}. The
galaxy is either part of the Cygnus\,A cluster or part of a cluster of
galaxies falling toward this cluster \citep{ledlow05}. The latter
scenario is supported by the fact that Cygnus\,A has a measured radial
velocity offset of 2197 km s$^{-1}$ from the mean cluster velocity
\citep{ledlow05}.

In the optical, broadened (FWHM 500-990 km s$^{-1}$) emission lines
\citep{tadhunter03}, as well as a blueshifted [O III]$\lambda$5007
line, are detected. The latter is identified as an outflowing, low
velocity wind, as is commonly observed in the UV and X-ray spectra of
Seyfert 1 galaxies.  \par In this paper we analyse three high spatial
resolution radio data sets at 5, 8 and 15 GHz, to study the jet and
counterjet in detail. In a companion paper \citep{steenbrugge08b} we
discuss the detection of a relic counterjet in a {\it Chandra} X-ray
image.  In a future paper we will discuss the relation between the
lobes and present a detailed study of the X-ray spectral properties of
the different brightness regions seen in the X-ray image. In the
present paper, the different radio observations are described in
Sect.~\ref{sect:obs}, in Sect.~\ref{sect:results} the observational
results are analysed, and interpretations are discussed in
Sect.~\ref{sect:disc}.

\section{Observations}\label{sect:obs}

In our analysis we used three archival radio data sets. The 5-GHz
radio data of Cygnus\,A (Fig.~\ref{fig:5_full}) were obtained by Chris
Carilli using the A to D VLA\footnote{ The Very Large Array is a
facility of the National Radio Astronomy Observatory, National Science
Foundation} configurations \citep{carilli91} and kindly given to us by
him. In addition, we use the 8\,GHz VLA A configuration data obtained
by Carilli and which were published by \cite{carilli96} and
\cite{perley96}.  The highest frequency band used was the
0.35$^{\prime\prime}$ resolution 15-GHz data made and published
by \cite{carilli96}, again using all four configurations of the
VLA. The data were reduced by him, using standard procedures in
AIPS as described in the above papers, and have undergone no further
processing by us.

\section{Results}\label{sect:results}

We first describe some of the details of the fine structure in the
radio images. A preliminary study of the trajectory of the jet
  and counterjet was reported by \cite{steenbrugge07}. Throughout
this paper when referring to Cygnus\,A we reserve the terms jet and
counterjet to refer to the western (approaching) and eastern
(receding) sides respectively.  

\subsection{Inner radio jet}\label{sect:inner}

\begin{figure}
\begin{center}
 \resizebox{\hsize}{!}{\includegraphics[angle=0]{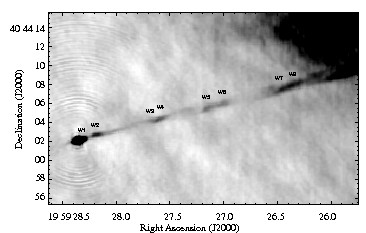}}
\caption{The inner jet of Cygnus\,A observed at 5~GHz. Note the weak
  emission between W1 and W2 which extends about half 
  way to W3. Thereafter, there there are only hints of smooth
  emission between the jet 
  knots, which become brighter from W7 onwards. The jet seems
  straight, but note the possible bend between W3 and W4. W7 and W8
  are the first jet knots in what we call the `outer
  jet'. \label{fig:5_wjet_det}} 
\end{center}
\end{figure}

\begin{figure}
\begin{center}
 \resizebox{\hsize}{!}{\includegraphics[angle=0]{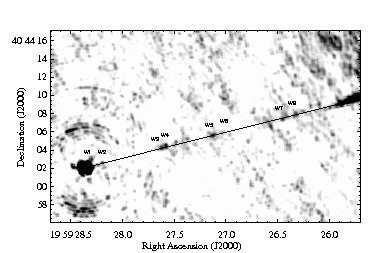}}
\caption{Detail of the inner jet of Cygnus\,A as observed at
  15\,GHz. The jet knots up to W6 are part of the inner jet. The inner
  edge of the 5-GHz lobe is just before W7. Note that the straight
  line drawn does not go through the centre of W4. The angle between
  the straight line drawn, the nucleus and the centre of W4 is
  1$^{\circ}$04$^{\prime}$. \label{fig:15_wjet_det}}
\end{center}
\end{figure}

\begin{figure}
\begin{center}
 \resizebox{\hsize}{!}{\includegraphics[angle=0]{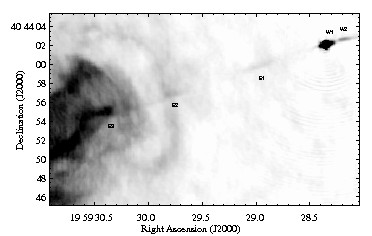}}
\caption{Detail of the inner counterjet of Cygnus\,A as observed at
  5\,GHz. Note how the jet knots are much weaker than on the jet side.
  The inner edge of the counterlobe does show very interesting
  features, such as the rather weak ``ring'' (which appears to
  surround E2) and the two ``antennae''(at E3). \label{fig:5_cjet}}
\end{center}
\end{figure}

We define the inner jet as the portion of the jet that is closer to
the core than the lobe is, i.e.\ where there
is less likely to be an interaction of the back-flowing lobe material with
the jet. We denote in Figs.~\ref{fig:5_wjet_det}
and~\ref{fig:15_wjet_det} W1 to W6 to represent the inner jet, and E1
and E2 in Fig.~\ref{fig:5_cjet} the inner counterjet, where E2 lies
close to the inner edge of the counterlobe. In the 5 and 8-GHz images
both the inner jet and the inner counterjet are observed. Both the jet
and counterjet consist of knots, with seemingly smooth emission
between some of the inner knots only detected in the 5-GHz image. In
the 15-GHz image only the bright knot at the inner edge of the lobe,
E2, is seen for the counterjet; however, for the jet there are 5 knots
interior to the 5-GHz lobe (which are labelled W1 to W5 in
Fig.~\ref{fig:15_wjet_det}). The inner jet and counterjet appear
straight, and have a maximum offset angle between knot W3 and W4 and the 
nucleus of 1$^\circ$04$^{\prime}$ $\pm$ 26$^{\prime}$. This possible
offset is best observed in the 15 GHz data
(Fig.\,\ref{fig:15_wjet_det}), in which the knots are more sharply
delineated, but is consistent with the 5- and 8-GHz images. No such
detailed information can be obtained for the inner counterjet, which
is significantly weaker in the 5- and 8-GHz images and not detected in
the 15-GHz image. The inner counterjet is at an angle of 182.79 $\pm$
1.5 degree to the western jet. The large error bar, compared to the
offset of the knots in the jet, is due to the much weaker emission
from the counterjet, which means only 2 knots are detected for the
counterjet in the 5- and 8-GHz images.

\subsection{Outer radio jets\label{sect:outer}}

\begin{figure}
\begin{center}
 \resizebox{\hsize}{!}{\includegraphics[angle=0]{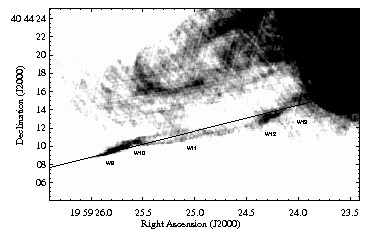}}
\caption{Detail of part of the outer jet near the western lobe,
  observed at 8\,GHz. The thin solid line indicates the direction of
  the inner jet. Note how the jet bifurcates between W10 and W11, a
  phenomenon known as threading. This figure clearly shows the bending
  of the jet between W9 and W12. \label{fig:8_threat}}
\end{center}
\end{figure}

\begin{figure}
\begin{center}
 \resizebox{\hsize}{!}{\includegraphics[angle=0]{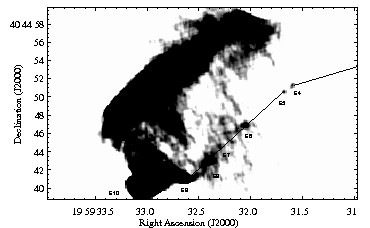}}
\caption{Detail of the outer counterjet (whose knots are labelled E4
  to E8) and the eastern hotspots (E9 and E10) observed at 15 GHz. Due
  to a lack of southern lobe emission in this band, we can follow the
  counterjet all the way to the hotspot region. The 2 solid lines
  indicate the direction of the counterjet, which changes direction
  between knots E4 and E5. \label{fig:cjet}}
\end{center}
\end{figure}

We define the outer jets as that portion of the jets that
clearly propagate through lobe plasma. The outer jet
starts at W7 (see Fig.\,\ref{fig:5_wjet_det}) and the counterjet is
represented by E3 to E8 (see Figs.\,\ref{fig:5_cjet} and
\ref{fig:cjet}).  In Cygnus\,A the western lobe is much more prominent
to the north of the jet; this is the case in all three radio bands
studied. This allows us to trace the jet well ``inside'' the
lobe. Also the eastern lobe is distributed asymmetrically around the
counterjet in the same sense, i.e. there is a preponderance of
  emission to the north (this is seen most clearly at 15-GHz,
see Fig.\,\ref{fig:cjet}), allowing us to observe the counterjet all
the way to a hotspot.  We observe that the jet bifurcates or
``threads'' just inside the inner edge of the lobe (between W10 and
W11 in Fig.~\ref{fig:8_threat}). Furthermore, the bends in the outer
jet are more easily observed than in the inner jet, and were first
noticed by \cite{carilli96}. Interestingly, the angle between W9 and
W12 (see Fig.~\ref{fig:8_threat}) is 1$^\circ$33$^{\prime}$ $\pm$
10$^{\prime}$, very similar to the angle determined between knot W3
and W4 in the inner jet. Further evidence that the jet bends is that
if we extend the inner jet to the hotspots (the extended jet line
shown in Fig.~\ref{fig:8_threat}), it does not terminate in either one
of the 2 brighter hotspots or even the third weaker hotspot. The
difference in angle between the extended jet line and a line from the
radio core to the middle of the inner bright hot spot is
$\sim$2$^\circ$30$^{\prime}$.

We do not observe the counterjet to thread in the same way the jet
does, but we do observe two bright knots which indicate that the
counterjet clearly bends (see Fig.~\ref{fig:cjet}). The angle
subtended at the nucleus by the 2 small knots E4 and E5 is only
40$^{\prime}$ $\pm$ 11$^{\prime}$. However, unlike the jet which seems
to bend around the extended jet line shown in
Fig.\,\ref{fig:8_threat}, these 2 knots seem to be placed where the
counterjet starts to bend through a large angle. The remaining knots
are only detected at 15\,GHz, as the lobe emission dominates in the
other bands.  The angle between the nucleus, and the E4 and E6 bright
knots is already 3$^\circ$33$^{\prime}$ $\pm$ 10$^{\prime}$, and
replacing E7 for E6 the angle is 6$^\circ$18$^{\prime}$ $\pm$
8$^{\prime}$. The jet bends by an angle of
$\sim$27$^\circ$31$^{\prime}$ between the nucleus, E4 and E9, the
weaker of the two hotspots on this side of the source. The width of
jet knot E8 in the 15-GHz image is $\sim$ 2$^{\prime\prime}$.9.

\begin{figure}
\begin{center}
 \resizebox{\hsize}{!}{\includegraphics[angle=0]{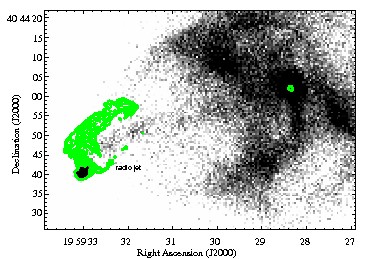}}
\caption{Detail of the {\it Chandra} 200 ks co-added ACIS-I image with a
  transfer function optimised to show the relic X-ray counterjet. The
  contours are from the 15\,GHz image, indicating the flux between
  0.008 and 0.05\,Jy. Note how the transversely extended X-ray
  counterjet goes through a low 
  luminosity radio region, seemingly pushing radio lobe plasma north.
  Furthermore, note that the radio counterjet is detected to the south
  of the X-ray counterjet. \label{fig:overlay}} 
\end{center}
\end{figure}

\subsection{The relic counterjet observed in X-rays}
\label{sec:relic}

In a companion paper (Steenbrugge, Blundell \& Duffy) we describe a
relic counterjet in Cygnus\,A which is revealed by deep {\it Chandra}
X-ray imaging.  In that paper, we discuss how its broad, transverse appearance (compared 
with the diameter of radio jets in this object) arises because of
expansion.  Fig.\,\ref{fig:overlay} shows where the
relic X-ray counterjet lies in relation to the most prominent northern 15-GHz 
lobe emission and to the current radio counterjet. We suggest that the
expanding relic counterjet is pushing the lobe and current
counterjet plasma radially outwards from its axis. This is a possible
explanation for the $\sim$27$^\circ$31$^{\prime}$ bend of the radio
counterjet, which is not observed for the relic X-ray counterjet. Also
the asymmetric luminosity of the counterlobe plasma, much brighter towards
the north, can be explained in this manner. 

\section{Discussion}\label{sect:disc}
\subsection{Precession}
\label{sect:precession}
In this section we consider precession of the jet axis as the origin
of the deviations from a perfect straight line as traced by the knots
forming the jet and counterjet of Cygnus\,A.  Precession, of course,
in its most general sense includes any change in the direction of the instantaneous 
spin axis.  All examples of precession are
fundamentally two-sided (applying to the emerging jet and counterjet
equally and simultaneously in the rest-frame of the nucleus).
Temporal variation in precession parameters, as long as two-sided and
instigated at the jet launch point by angular momentum changes (e.g.\
caused by variation in the fuelling), are properly described as
precession and are distinct from Scheuer's Dentist's Drill phenomenon
\citep{scheuer82}. 
This phenomenon is a response of a jet, on {\em one side} of a
source, to local 
conditions (for example, buoyancy effects corresponding to local
motions or inhomogeneities); we note that dentists' drills are
one-sided. Although the dentists' drill could work for both sides
of the jet, as the effect is environment dependent, one would not
expect a symmetric pattern.

Here we investigate whether sustained periodic precession is a
plausible explanation for the curvature seen in Cygnus\,A's jets.
Sustained periodic precession is famously exhibited by the jet axis of
the Galactic microquasar SS\,433, which has a relatively rapid
precession period of $\sim 162$ days.  A geometric model describing
the precession of the jet axis in this, or any similarly behaving
object, is presented by \citet{hjellming81}.  In this model, symmetric
jets are launched at speed $\beta$ along an axis which traces a cone
throughout a precession period $P$, of opening angle $\psi$ inclined
to our line of sight at angle $i$, in terms of a precession phase
$\phi$.  The jet material, once launched, is assumed move
ballistically, i.e.\ suffer no acceleration or deceleration.

Reasonable fits to the curvature of the jets in Cygnus\,A can be
obtained with the following parameter values: a cone opening angle of
1$^{\circ}$ $\pm$ 0.30$^{\circ}$, an angle between this cone axis and
our line of sight of 
60$^{\circ}$ $\pm$ 5$^{\circ}$, and an angle of the cone axis
projected on the plane of the 
sky with respect to the East-West line of 14.7$^{\circ}$ $\pm$
1$^{\circ}$.  $D$ is determined 
by the redshift 0.05607 for the cosmology assumed in this paper, and
is not considered to be a free parameter.  The characteristic angular
periodicity of the appearance on the sky of a precessing jet is
determined by $\beta P/D$ (where $D$ is the distance of the precessing
jet from Earth).  Different values of $P$ and $D$ will cause a simple
magnifying effect on this angular scale, but higher values of $\beta$
will introduce asymmetric curvature on the opposite sides due to increasing asymmetric light-travel time effects.  

\begin{figure}
\begin{center}
 \resizebox{\hsize}{!}{\includegraphics[angle=0]{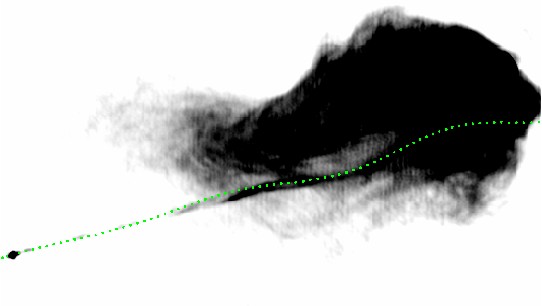}}
\caption{5-GHz image of the jet side of Cygnus\,A, with the best fit
  precession model as fitted to 15~GHz overlaid as green dots. The
  jet knots are launched anti-parallel along a jet axis which
  precesses, tracing out a cone of opening angle 1$^{\circ}$. Details
  of parameters for the precession model fitted are listed in the text.
  \label{fig:precession}}
\end{center}
\end{figure}

\begin{figure}
\begin{center}
 \resizebox{\hsize}{!}{\includegraphics[angle=0]{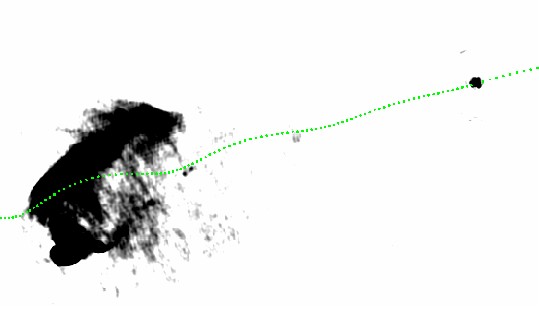}}
\caption{15-GHz image of the counterjet side of Cygnus\,A,
  with the best fit precession model overlaid as green dots. Note that 
  the expansion of the relic counterjet (shown in
  Fig.\,\ref{fig:overlay}) will push the counterjet to the
  south. Therefore, precession does not explain the
  $\sim$27$^\circ$31$^{\prime}$ 
  bend in the counterjet. 
  \label{fig:precession_cj}}
\end{center}
\end{figure}

The projection of a precessing jet with these parameters for
particular related values of $\beta$ and $P$ are shown for the
5~GHz jet in Fig.\,\ref{fig:precession} and for the 15~GHz counterjet
in Fig.\,\ref{fig:precession_cj}. The fit was made to the 15~GHz image
and has values: $\beta$ = 0.35 c, $P$ = 10$^8$ days (or 0.274 Myr),
a line of sight angle of 60$^{\circ}$~and a cone opening angle of
1$^{\circ}$. The allowed pairs of $\beta$ and $P$ that give a
decent fit range between: $\beta = 0.3\,c$ and $P$ = $1.17 \times
10^8$ days to $\beta = 0.5\,c$ and $P$ = $0.7 \times 10^8$ days. To
quantify the goodness of the precession model we traced the jet by-eye
and calculated the standard deviation between the precession model fit
and the by-eye fit. For both Right Ascension and Declination we use proper
arcseconds. We then vector added those standard deviations and
obtained a difference between our best fit precession model and the by
eye fit of 0.072$^{\prime\prime}$ (proper arcsec). For a fit with
$\beta$ = 0.5 this value worsens to 0.080$^{\prime\prime}$, and for
$\beta$ = 0.25 this value is 0.086$^{\prime\prime}$. An opening angle
of 1.4$^{\circ}$ worsens this value to 0.079$^{\prime\prime}$. A line
of sight angle of 65$^{\circ}$~worsens the goodness of fit to
0.075$^{\prime\prime}$. Finally, assuming no precession (i.e.\ a
completely straight jet), the fit worsens to 0.103$^{\prime\prime}$. 

\begin{figure}
\begin{center}
 \resizebox{\hsize}{!}{\includegraphics[angle=0]{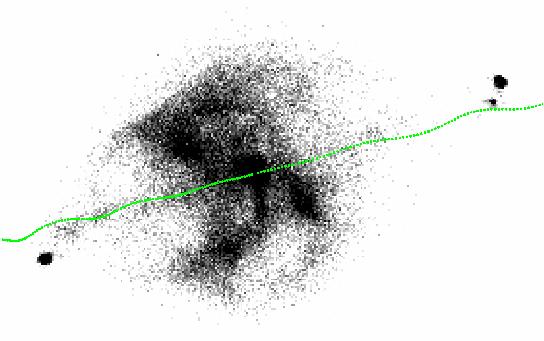}}
\caption{The 200~ks image of Cygnus\,A (Steenbrugge, Blundell \& Duffy 
  companion paper), with the precession model fitted to the 15-GHz
  image and shown in Figs. \ref{fig:precession} and
  \ref{fig:precession_cj} overlaid as green dots. 
  \label{fig:x-ray_precession}}
\end{center}
\end{figure}

Thus, for the approaching jet shown in
Fig.\,\ref{fig:precession}, a precessing jet axis gives encouraging results.  It
is particularly interesting that the range of jet speeds which work
are consistent with the slower jet speeds inferred from an independent
method and literature described in Sec\,\ref{sect:jetspeed} and \ref{sec:literature}. The fit to the
counterjet is clearly poorer than the fit to the jet. However,
allowing for a $\sim$~1$^{\circ}$ difference from the assumed
180$^{\circ}$ angle between the jets does lead to a much improved
counterjet fit. The considerable change in angle
($\sim$27.5$^{\circ}$) of the current radio counterjet seen in the
east of Cygnus\,A occurring between E4 and E5 is unlikely to be
due to precession, but could be influenced by the expansion of the
relic counterjet (see Section\,\ref{sec:relic}). Actually, the
precession fit obtained from fitting the 15-GHz image goes
surprisingly well through most of the relic X-ray counterjet
(Steenbrugge, Blundell \& Duffy, companion paper), as shown in
Fig.~\ref{fig:x-ray_precession}. Although the fit is poor at the outer
part of the relic X-ray counterjet, the decent fit for part of the jet
suggests that precession has been fairly constant for different jet
episodes, considering that the precession phase likely was different.

We remark that if the jet knots move at speeds $\sim 0.3\,c$, a rather
low value for the jet speed, that the
time taken to reach the current location of the hotspots is $\sim 10^6$
years. 

\subsection{Other models for jet bending}
We will briefly review some of the other models that in the past have
been proposed to explain bending of jets or tails and
suggest why they are unlikely to apply to Cygnus~A. 

\par A possible mechanism for jet bending, that can easily be excluded
for Cygnus~A, is the bending due to the orbiting of the black hole within
its galaxy. For an orbiting object the jet curvature is
plane symmetric, thus the bend in the jet and counterjet are towards
the same side.  Whereas in precession, the jets are point symmetric (at least
for slow speeds), i.e. the bends have a 180$^{\circ}$ mirror on the
other side.  The curvature of Cygnus\,A's jets is consistent with
point symmetry (see Fig.\,\ref{fig:precession} and \ref{fig:precession_cj}). 

\par \citet{eilek84}, studying the wide-angle tailed source 3C\,465,
tried several models to explain the C-symmetric, i.e.\,mirror
symmetric, bending of its outer jets. As the jets of 3C\,465 inside
the host galaxy are straight, all models ascribe the effect to the
intra-cluster medium (ICM). Although no model appeared convincing to
those authors, the most likely model is that the bending is due to the
large-scale velocity of magnetic fields in the ICM. 3C\,465 is in the
centre of a cluster, therefore it likely has no large-scale velocity
of its own to explain the bending. The bending in the outer jet of
Cygnus\,A is outside the host galaxy, and Cygnus\,A is either part of
a group or a cluster. However, large-scale velocity or magnetic field
structure would also bend the lobe, and this is not observed.  The
fact that in Cygnus\,A the bending reverses (first away from the
straight line in Fig.\,\ref{fig:8_threat} and then back again), would
require highly ordered fields on rather small scale and with a very
specific configuration. Finally, this explanation cannot explain the
subtle bending of the inner jet.

\par A third model proposed by \cite{wirth82} is that the
gravitational potential of a companion galaxy torques and thereby
bends the jet.  \citet{wirth82} use this model to explain the jet shape
in the dumbbell galaxy NGC\,326, which \cite{gower82} modelled as
precession. \cite{wirth82} assume that in NGC~326 only the inner part
of the jet is a current jet, and that the more extended emission is
from a previous jet episode. This is an unlikely explanation for the
bending of the jets in Cygnus\,A, as the galaxy, although in either a
group or a cluster, does not have a very close companion of similar
mass like the dumbbell galaxy of NGC\,326. Furthermore, the scale over
which the jet bends is such that the companion should have a rotation
period of less than the inferred precession period (of the order $3
\times 10^5$ years), for the range of jet speeds inferred in
Sect.\,\ref{sect:precession}. For a companion galaxy at 20\,kpc
distance this would require an orbital speed close 0.1\,$c$,  an
implausibly high value.

\par Finally, it is possible that gas pressure within the host
galaxy itself, due to the host galaxy rotation, bends the jet. As the
rotation speed is a function of radius in the galaxy, a very
efficient, i.e. near instantaneous, transfer of angular momentum of
the gas to the jet, could cause the jet to first bend in one direction
and then turn back again. Namely, the jet would be rotating too fast
for the radius at which it finds itself after travelling further
out. In this model, once the jet leaves the host galaxy, or the gas
pressure is too low, the jet would travel in a straight line. To
explain the bend in the outer jet, which is well outside the host
galaxy, one needs to postulate that the jet speed changed over
time. As a result the amount of bending, and thus the angle from
which the jet left the host galaxy varies with time. Although
this remains to be explored in detail, an argument against this
possibility is that the angle over which the inner and outer jet bend
is (within errors) the same, which would be unlikely if the bending is
caused by two different processes.

\subsection{Comparison of precession period with other sources}

\begin{figure}
\begin{center}
 \resizebox{\hsize}{!}{\includegraphics[angle=-90]{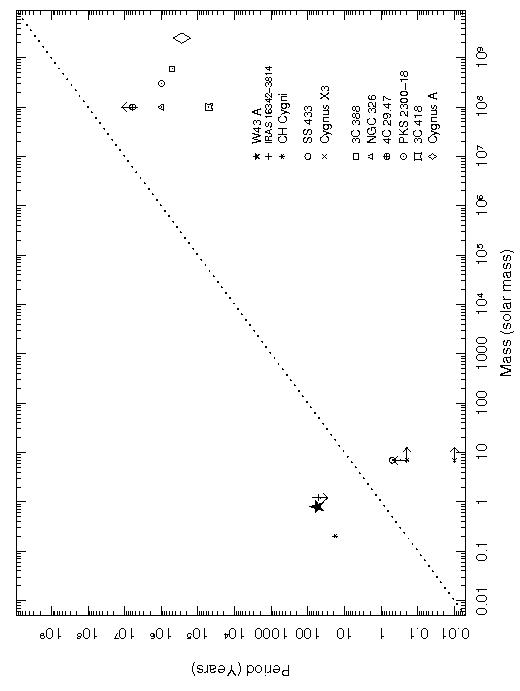}}
\caption{\label{fig:comp} The plot shows the precession period and
  (estimated) mass for young stellar objects,
  microquasars and AGN showing evidence of precessing jets. The mass
  for those sources 
  without a determined mass quoted in the literature was assumed as
  follows: for the young stellar objects 1M$_{\odot}$, for SS\,433 and
  Cygnus\,X-1: 7M$_{\odot}$ and AGN 10$^8$ M$_{\odot}$. References:
  W\,43A: \citet{vlemmings06}, IRAS\,16342$-$3814: \citet{sahai05},
  CH\,Cygni: \citet{crocker02}, SS\,433: \citet{blundell04}, Cygnus~X-3:
  long period and mass: \citet{mioduszewski01}, short period:
  \citet{miller-jones04b}, 3C\,388 and NGC\,326: \citet{gower82} for
  period, \citet{woo02} for the mass of the former, 4C\,29.47:
  \citet{condon84}, PKS\,2300$-$18: \citet{hunstead84} for the period
  and \citet{wu04} for the mass, 3C\,418: \citet{muxlow84}, Cygnus~A:
  \citet{tadhunter03} for the mass, see this paper for the period. For
  Cygnus~A the 
  size of the symbol corresponds to the error bars in mass and period.}
\end{center}
\end{figure}

In this section we compare the results from our precession model
(Sec.\,\ref{sect:precession}) with the results for precessing jets
that are available in the literature.  For most objects the mass is
not yet determined. Therefore we deemed that the young 
stellar objects have a mass of 1M$_{\odot}$, the microquasars of
7M$_{\odot}$ and the AGN of 10$^8$ M$_{\odot}$, unless there was a
value in the literature. (Considering the spread in mass for the
objects considered, the uncertainty in mass is not too problematic.)
Fig.~\ref{fig:comp} shows the relationship between mass and precession
period for younger stellar objects, microquasars and AGN.

AGN clearly have longer precession periods than either microquasars or
young stellar objects.  Considering that microquasars are likely to be
much better analogues of quasars then young stellar objects (e.g.\
because of the similarity of their jet speeds; the jet speeds of young
stellar objects are only of order 100\,km\,s$^{-1}$), it is
interesting to note that the distribution of AGN points and that of
the microquasars lie a similar distance below the $y = x$ line,
drawn solely to guide the eye, and not a fit to the data. This may
be an indication that precession periods in these objects increase
with mass.  

This possible correlation would disappear if additional
  data-points are located in either the bottom right or the top left
  corner of Fig.~\ref{fig:comp}.  The first of these would occur for
  AGN having very short precession periods.  Fine angular resolution
  is necessary to be able to detect very short precession periods in
  distant AGN jets. There have been claims
  in the literature of very short precession periods in AGN, mostly
  from VLBI observations. For example, the 30.8-yr derived precession
  period in PKS\,1830$-$211, which is a blazar with two nuclei
  \citep{nair05}.  However, these short precession periods are most
  likely due to the orbital motion of the binary as explained and
  modelled by \cite{kaastra92} and recently reviewed by
  \cite{lobanov05}. The difference in precession periods for
  microquasars and AGN should be confirmed by more data, especially
  better VLBI data to rule out precession of AGN on timescales of tens
  of years or less. The second of these would occur if Galactic
  microquasars are discovered to have very long precession periods.
  In general, the data available for young stellar objects and
  microquasars are not sensitive to precession periods longer than
  about a 1000 years.  At present, we only know of one microquasar,
  SS433 embedded within the W50 nebula, with precessional information
  which can be traced back to $\sim 10^5$ years ago.

Interestingly, the microquasars have shorter precession periods ($\sim
162$ days in the case of SS\,433) compared to the young stellar
objects ($\sim 50$ years), although they are certainly more massive
than the young stellar objects. (The uncertainty in the mass does not
undermine this finding.)  It is clear that more data points, and
including neutron stars with precessing jets, would be very helpful in
aiding our understanding of jet precession and the interpretation of
Fig.~\ref{fig:comp}.

\subsection{Causes of precession}
In the following, we detail three possible mechanisms for the
precession observed in the jet. For all cases we assume that the jet
is launched from the inner part of the accretion disc, but outside the
warping radius. The Lense-Thirring effect will realign the angular
momentum of the accretion disc within the warping radius, with that of
the central black hole.  All three mechanisms described involve
creating a misalignment of the accretion disc, with either the black
hole or the galactic stellar disc. Therefore the secure
  identification of precession in
active galactic nuclei with a super-massive black hole would seem to
indicate that the jet must be launched from the accretion disc and not
the black hole.

\subsubsection{Gas flow onto accretion disc}
Gas, perhaps from a recent merger, channelling onto the accretion disc
can cause a misalignment of the angular momentum vector of the
accretion disc and the black hole. Redshifted HI\,21-cm absorption
\citep{conway95} and redshifted H$_{2}$ emission lines
\citep{bellamy04} are observed in the nucleus of Cygnus~A, most likely
indicating 
gas inflow. Therefore this is a possible mechanism for Cygnus~A. The
realignment of the accretion disc with the black hole was studied by
\cite{scheuer96} for small angle misalignment. In the scenario they
studied, Lense-Thirring precession is taken into account causing the
accretion disc to warp as the black hole forces a realignment of the
innermost part of the accretion disc. The warping radius is the radius
where the Lense-Thirring effect becomes negligible. The realignment
occurs within 3$\times$10$^8$ years for a 10$^9$ M$_{\odot}$ black
hole, for an accretion rate of 1 M$_{\odot}$ yr$^{-1}$ and a warping
radius of 100 Schwartzschild radii. This timescale is probably a good
estimate for Cygnus\,A. Considering that the age for the jet that we
derive from the precession model is much less than 10$^8$ years, this
model is certainly possible from the observations.

\subsubsection{Companion black hole}
Another possible mechanism for precession is a companion black
hole. As the companion black hole will exert a torque on the
accretion disc, and thereby misalign it with the primary black hole
spin, the resulting precession is similar as described in the above
section. This mechanism is used to explain precession in microquasar jets
and pre-main sequence star jets, albeit with the companion then being
an unknown star instead of a black hole. \cite{canalizo03} did detect
a possible second nucleus with the Keck, at a distance of
0.4$^{\prime\prime}$ or $\sim$ 400 pc from the jet ejecting or primary
black hole. Considering that there is a molecular cloud \citep{bellamy04}
and a possible companion nucleus \citep{canalizo03}, both causes
(infalling gas which causes the accretion disc to be misaligned and a
companion black hole that causes the accretion disc to be misaligned)
are possibly at work in Cygnus\,A.
 
\subsubsection{Angular momentum disc and host galaxy}
A third possible mechanism for driving the precession is an offset in
angular momentum vector direction between the angular momentum in the
accretion disc and the angular momentum of the gas in the host
galaxy. In this case the gas in the host galaxy will torque the
accretion disc, thereby causing precession, as well as causing a
misalignment of the accretion disc angular momentum and that of the
super-massive black hole.

\subsection{Independent estimates of jet speeds: from jet knots distances}\label{sect:jetspeed}

Under certain assumptions one can calculate the jet speed from the
observed distances from the core for the knots in the jet and
comparing these with the distances for the knots in the counterjet. A
first assumption is that corresponding knots on the jet and counterjet
side were emitted at the same time, and therefore are not shocks
  along the path of the jet due to environmental differences. The
  assumption is minimised for the inner jets, however, as the the
  emission from the inner jet is very weak, it is not 
  possible to derive a reliable jet speed just from the inner jet. The second
assumption is that the 
knots were emitted with the same speed in anti-parallel directions,
and have a similar deceleration/acceleration along their paths; i.e.\
that there are no environmental differences between the two sides.  To
be able to measure these distances, i.e.\ take out the effect of
projection, one needs to know the inclination angle of the jet with
respect to our line of sight, $\theta$. We find this angle to be 
60$^{\circ}$ from our precession modelling (Sec.\,\ref{sect:precession}).
Finally there is a more practical problem, the knots on the
counterjet side are much weaker, hence we do not detect as many
knots on the counterjet side as we do on the jet side. This leads to
some uncertainty about how the knots should be paired. Bearing all
these caveats in mind we find the jet speed, $\beta$, using:

\begin{equation}
\beta \cos\theta = \left( {x-1}\over {x+1} \right)
\end{equation}
where $x$ is the distance between the core and a knot in the jet
divided by the distance between the core and corresponding knot in the
counterjet.

Table~\ref{tab:knotspeed} lists the jet speeds for those knots that we
assessed as being plausible pairs. The extreme (but formally allowed)
values for jet speed for any pairing were 0.98\,$c$ (knot W7 and E1)
and 0.07\,$c$ (knots W12 and E6). Note that several jet knots can be
paired plausibly with more than one counterjet knot, and indeed,
several counterjet knots can be paired with more than one jet
knot. This gives a clear indication of the uncertainty in pairing the
knots. The inference from Table\,1 is that the jet speeds are
fairly slow, lower than the best fit jet speed derived from our
  precession model.

\begin{table}
\begin{center}
\caption{Jet speeds derived from plausible matchings of different
  pairs of knots 
  from the jet and counterjet. As fewer knots are observed in the
  counterjet there is no unique pairing. We use an inclination
  angle of 60$^{\circ}$ in 
  calculating the jet speed. The labels for the knots correspond to
  those in Figs.~\ref{fig:5_wjet_det} to \ref{fig:cjet}.} 
\label{tab:knotspeed}
\begin{tabular}{c|c|l}\hline \hline
jet knot  & counterjet knot    & $\beta$ \\\hline
W4        & E1                 & 0.16   \\
W7        & E2                 & 0.27   \\
W9        & E3                 & 0.21   \\
W10       & E3                 & 0.26   \\
W12       & E4                 & 0.22   \\
W12       & E5                 & 0.19   \\
\end{tabular}
\end{center}
\end{table}

\subsection{Jet speeds for Cygnus\,A from the literature}
\label{sec:literature}
\cite{krichbaum98} use two epochs of 22-GHz VLBI data to determine the
jet speed for Cygnus\,A from the proper motion of the knots observed,
as well as from the difference in luminosity between the jet and
counterjet. From the observed proper motion of knots in the jet they
derive an apparent jet speed of less than 0.27\,$c$ (recalculated for
the Hubble constant we assume throughout this paper) and an
inclination angle of less than 50$^{\circ}$. \cite{krichbaum98}
determined the jet-to-counterjet ratio near the nucleus to be 1.3
although the steeper spectrum of the pc-scale counterjet compared with
the pc-scale jet suggests that this side is being absorbed so this
number is an upper limit and may be closer to unity. 
Using this flux ratio between the jet and the counterjet, 
and not taking light-travel time effects into account, they derive jet speeds
between 0.27\,$c$ and 0.82\,$c$.  However, \cite{krichbaum98} assert
that the low jet speeds 'are too low to be considered realistic for
Cygnus A'. \cite{sorathia96} used the same methods with VLBA and VLBI
observations to derive a jet speed between
0.5\,$c$ and 0.8\,$c$ for an inclination angle to the line of sight
between 45$^{\circ}$ and 75$^{\circ}$, again assuming a Hubble
parameter of 73\,km\,s$^{-1}$ Mpc$^{-1}$.  

\cite{carilli96} measured the flux ratio to be between 5 and 21 for
the pc-scale jet  (using VLBI data) and 2.6 $\pm$ 1 for the
kpc-scale jet, but don't derive a jet speed for the kpc jet. However,
\cite{pelletier86} derive an upper limit to the jet speed of 0.6c in
Cygnus A, using hydrodynamical 
  modelling of the jets and the shock conditions in the hotspots. 

Therefore, the range of jet speeds we derive for the kpc-scale jet,
which has potentially undergone deceleration, is not unrealistic,
compared to previously measured jet speed for the pc-scale jet and the
modelled value for the kpc-scale jet.

\subsection{Comparison of jet speed with other precessing AGN}

In the literature compilation of precessing sources plotted in
Fig.~\ref{fig:comp} there were 5 AGN. For 4 of these the authors gave
the jet speed (or upper limit) derived from fitting a precessing jet
model. These jet speeds are listed in
Table~\ref{tab:jetspeed}. Interestingly, in all 4 cases the jet speed
was constrained to be less than 0.3c. Therefore, it seems reasonable
that we find a jet speed of similar magnitude in Cygnus\,A. 

\begin{table}
\begin{center}
\caption{The table lists the literature values for jet speeds derived
  assuming a precession model and the relevant reference. } 
\label{tab:jetspeed}
\begin{tabular}{l|l|l}\hline \hline
source      & jet speed $\beta$ & reference \\\hline
3C 388      & 0.15             & \cite{gower82} \\
NGC 326     & 0.20             & \cite{gower82} \\
4C29.47     & $\leq$0.2        & \cite{condon84} \\
PKS 2300-18 & $\leq$0.3        & \cite{hunstead84} \\\hline
\end{tabular}
\end{center}
\end{table}
 
\subsection{Wardle \& Aaron's study}

\citet{wardle97} studied the jet to counterjet luminosity ratio for
the 13 3CR sources with deep VLA images made by \citet{bridle94}. To
exclude possible effects from interaction with lobe material, they
studied only the inner, straight kpc-scale jets.  Wardle \& Aaron
assumed a spectral index of 0.6, and that the flux density enhancement
is $D^{\delta}$ with $D$ the Doppler factor and $\delta = 2.6$
appropriate for a continuous emission model of the
jet. \citet{wardle97} conclude from the measured luminosity ratio that
the jet speeds for the kpc-scale jets are between 0.6\,$c$$-$0.7\,$c$,
and that the intrinsic asymmetry in the jets is small.

The analysis by \citet{wardle97} is statistical, as the inclination
angles are unknown, and they admit that their sample is biased towards
the largest sources. Considering the statistical analysis is over a
biased sample, we see no reason to shy away from considering slow jet
speeds for the kpc-scale jet in Cygnus\,A.

\section{Conclusions}

We have examined the detailed structure of the radio jet and
counterjet in multi-band, multi-VLA configuration radio images kindly
supplied by C.\ Carilli.

The trace of the radio knots that delineate the jet and counterjet
deviates from a straight line, at least part of which can be
satisfactorily fitted with the precession model of
\citet{hjellming81}. We find plausible parameter values for the
precession model fits with jet speeds \lesssim 0.5\,$c$, which are
consistent with  independent methods of estimating the jet speed in
Cygnus\,A.

The fact we observe precession in AGN seems to indicate that the jet
is launched from the accretion disc. Comparing precession periods
for sources with different masses, we find that young stellar
objects have longer periods than microquasars, but that AGN have the
longest periods.

\section*{Acknowledgements}
KCS would like to thank St John's College, Oxford for a fellowship and
John Everett for several discussions relating to the work presented in
this paper. KMB expresses her gratitude to the Royal Society and both
authors thank especially Chris Carilli for providing the radio data
analysed in this paper and Robert Laing for helpful discussion. We are
very grateful to the referee for helpful comments.

\bibliography{references}

\label{lastpage}

\end{document}